\documentstyle[12pt]{article}
\makeatletter
\def\@maketitle{\newpage
 \null
 {\normalsize \tt \begin{flushright}
  \begin{tabular}[t]{l} \@date
  \end{tabular}
 \end{flushright}}
 \begin{center}
 \vskip 2em
 {\LARGE \@title \par} \vskip 1.5em {\large \lineskip .5em
 \begin{tabular}[t]{c}\@author
 \end{tabular}\par}
 \end{center}
 \par
 \vskip 1.5em}
\makeatother
\begin{document}
\setlength{\baselineskip}{16pt}
\title{2D Quantum Gravity \\ in the Proper-Time Gauge}
\author{
        Ryuichi NAKAYAMA \thanks{
                          nakayama@particle.phys.hokudai.ac.jp}
\\[1cm]
{\small
    Department of Physics, Faculty of Science,} \\
{\small
           Hokkaido University, Sapporo 060, Japan}
}
\date{
  EPHOU-93004  \\
  December 1993
}

\maketitle

\begin{abstract}

A two-loop (cylinder) amplitude of the 2d pure gravity theory is obtained  in
the proper-time gauge ($g_{00}=1$, $g_{01}=g_{10}=0$) in the continuum
formulation.  The constraint $T_{01}=0$ is solved and used to reduce
the problem of field theory to that of quantum mechanics.
This reduction can also be proved by using a conformal Ward identity.
The amplitude depends on the lengths
$l_1, l_2$ of the boundaries, the proper time $T$ and a non-negative integer
$m$ associated with winding modes around the boundaries.
\end{abstract}

\newpage

\newcommand {\n}{\nonumber\\}
\newcommand {\nn} {\nonumber}
\newcommand {\cleqn}{\setcounter{equation}{0}}
\renewcommand {\theequation}{\arabic{equation}}
\newcommand {\beq}{\begin{equation}}
\newcommand {\eeq}{\end{equation}}
\newcommand {\beqa}{\begin{eqnarray}}
\newcommand {\eeqa} {\end{eqnarray}}
\newcommand {\beqc}{\beq\begin{array}{c}}
\newcommand {\eeqc}[1]{\label{#1}\end{array}\eeq}
\newcommand {\beql}{\beq\begin{array}{l}}
\newcommand {\eeql}[1]{\label{#1}\end{array}\eeq}
\newcommand {\lfrac}[2]{\frac{\displaystyle #1}{\displaystyle #2}}
\newcommand {\sfrac}[2]{\displaystyle \frac{#1}{#2}}
\newcommand \han{\frac{1}{2}}


Much progress in the formulation of 2d quantum gravity has been made since
the equivalence of 2d quantum gravity and
dynamical triangulation (DT)\cite{kkm} was established.
\cite{matrix}~$-$~\cite{Ishi}
As for the problem of understanding 2d gravity solely in the continuum
theory, there still remain several technical difficulties, although
there are some important developments \cite{dkd}~$-$~\cite{GN}.
Especially, loop amplitudes have not been derived by the
quantization of the gravitationally induced action in spite of some attempts
\cite{Moore} \cite{Seiberg}.

In this letter a two-loop (cylinder) amplitude of the pure gravity theory will
be derived in the continuum formulation of 2d gravity.  For this purpose we
should first specify a suitable gauge.  A natural coordinate system
on a cylinder is obtained by choosing its height as the direction of the time
coordinate $x^0$ and the cycle perpendicular to it as that of the space
coordinate $x^1$.  In this coordinate frame we will make the following gauge
fixing for the metric (proper-time gauge \cite{proper}):
\beq
g_{00}=1,  \quad \quad g_{01}=g_{10}=0 .
\label{1}
\eeq
The component $g_{11}$ is denoted as $\gamma (x^0,x^1)$.

Let us first derive a conformal Ward identity.  We start from the proper-time
gauge (\ref{1}) and perform an infinitesimal reparametrization of $x^0$,
$\delta x^0= \epsilon (x^0)$.  Then the metric is no longer of the form of
(\ref{1}).  To return to (\ref{1}) we should rescale the metric as
$g_{\mu \nu} \rightarrow (1+2 \epsilon' ) g_{\mu \nu} $.
The total change of $\sqrt{\gamma}$ is $\delta \sqrt{\gamma}=
\epsilon ' \sqrt{\gamma}-\epsilon \partial_0 \sqrt{\gamma}$.
Since the action and the integration measure are invariant under
reparametrizations but not under Weyl rescalings, we obtain the following
Ward identity
\beqa
& & < \int d^2x \epsilon '(x^0) \sqrt{g(x)} (c \, R(x)+ \lambda) \sqrt{
\gamma (x_1)} \cdots \sqrt{\gamma (x_n)} >  \n
&=& \sum_{j=1}^n [\, -\epsilon (x^0_j) \partial / \partial x^0_j +
\epsilon ' (x^0_j) \, ]< \sqrt{\gamma (x_1)} \cdots \sqrt{\gamma (x_n)} >,
\label{WI1}
\eeqa
where $\sqrt{g} R = -2 \partial_0^2 \sqrt{\gamma}$, $c$ is
 a constant proportional to the conformal anomaly and $\lambda$ a cosmological
constant.  By removing $\epsilon$ from (\ref{WI1}), integrating over $x^1_j$
and defining a length of the loop at time $x^0$ by
$l(x^0)=\int^{\pi}_0 dx^1 \sqrt{\gamma (x)}$, we have
\beqa
& & [ \, -2c (\partial /\partial x^0)^3 + \lambda \partial /\partial x^0 \, ]
   < l(x^0) l(x^0_1) \cdots l(x^0_n) > \n
&=& \sum^n_{j=1} [ \, \delta(x^0-x^0_j) \partial / \partial x^0_j +
    \delta ' (x^0 - x^0_j) \, ] < l(x^0) l(x^0_1) \cdots l(x^0_n)>.
\label{WI2}
\eeqa
This is a closed equation for Green functions of $l(x^0)$.  Existence of such
an identity suggests that there should be a quantum-mechanical action for
$l(x^0)$.   Indeed there is such an action and it can be written as
\beq
S_l = \int dx^0 [ \, \frac{c}{l}(\frac{d}{dx^0} l)^2+\frac{\lambda}{2}l+
     \frac{a}{l} \, ],
\label{S}
\eeq
where $a$ is some constant whose value is unknown at present.
It is easy to derive
(\ref{WI2}) from the action (\ref{S}) by a change of variables
$l(x^0) \rightarrow l(x^0) +\epsilon '(x^0) l(x^0)-\epsilon(x^0) l'(x^0)$.
One of the purposes of this letter is to derive  (\ref{S}) from the
gravitationally induced action.  It will be shown that the constant $a$ takes
discrete values.

As is well-known, the continuum action in 2d gravity is given by the induced
action which is obtained by integrating out the matter fields coupled to
gravity.  In the conformal gauge $g_{\mu \nu}=e^{\phi} \hat{g}_{\mu \nu}$,
the induced action coincides with Liouville action
\beq
S_L (\phi ; \hat{g}_{\mu \nu}) = \int d^2x \sqrt{\hat{g}} [ \lfrac{1}{4}
\hat{g}^{\mu \nu} \partial_{\mu} \phi \partial_{\nu} \phi
+ \lfrac{1}{2} \phi \hat{R} +4\lambda e^{\phi}],
\label{2}
\eeq
where $\phi$ is the conformal mode, $\hat{g}_{\mu \nu}$ is the reference
metric and $\lambda$ is a cosmological constant.  Although $S_L$ is local,
in gauges such as (\ref{1})  the induced action $S_{ind}$ is non-local.
Let us parametrize an arbitrary metric $g_{\mu \nu}$ as follows:
\beq
g_{\mu \nu} dx^{\mu} dx^{\nu}=e^{\phi}\hat{g}_{\mu \nu}dx^{\mu}dx^{\nu}
=e^{\phi}(dz+\mu d\bar{z})(d\bar{z}+\bar{\mu}dz),
\label{3}
\eeq
where $z=x^0+ix^1, \bar{z}=x^0-ix^1$.  We introduce two functions $f(x^0,x^1)$
and $\bar{f}(x^0,x^1)$ to define $\mu$ and $\bar{\mu}$:
\beq
\mu=\bar{\partial}f/\partial f \quad \quad \mbox{and} \quad \quad
\bar{\mu}=\partial \bar{f}/\bar{\partial} \bar{f},
\label{4}
\eeq
where $\partial=(\partial_0-i\partial_1)/2$, $\bar{\partial}=(\partial_0
+i\partial_1)/2$ and $\partial_\mu=\partial/\partial x^{\mu}$.
The induced action $S_{ind}$ is then given \cite{Verlinde} by
\beq
S_{ind} = S_L(\phi;\hat{g}_{\mu \nu})+S_f(f)+\bar{S_f}(\bar{f})+S_{l.c.}(\mu,
\bar{\mu}),
\label{5}
\eeq
where $S_f(f)$ is Polyakov's light-cone gauge action \cite{Polyakov}:
\beq
S_f(f)=\lfrac{1}{2} \int d^2x [ \frac{\partial^2f \partial \bar{\partial}f}
{(\partial f)^2}
      - \frac{\partial^3f \bar{\partial}f}{(\partial f)^2} ].
\label{6}
\eeq
Here $\bar{S_f}(\bar{f})$ is the complex conjugate of $S_{f}(f)$.  The local
counterterms are determined by the requirement of general coordinate
invariance and we have
\beq
S_{l.c.}(\mu,\bar{\mu})= \int d^2x (1-\mu \bar{\mu})^{-1} \{ 2\partial \mu
\bar{\partial} \bar{\mu}-\mu(\bar{\partial}\bar{\mu})^2-\bar{\mu}(\partial
\mu)^2 \} .
\label{7}
\eeq
The action (\ref{5}) is indeed invariant under an infinitesimal coordinate
transformation ($\delta z=\epsilon(z,\bar{z})$)
\footnote{Here $\delta \bar{z} = \bar{\epsilon}(z,\bar{z})$, which is a
complex conjugate of $\epsilon(z,\bar{z})$, is formally set to zero.
It is a straightforward task to recover $\bar{\epsilon}$.  }
\beqa
\delta \phi &=& \epsilon \partial \phi+(\partial +\bar{\mu}\bar{\partial})
\epsilon, \n
\delta \mu &=& \epsilon \partial \mu+(\bar{\partial}-\mu\partial)\epsilon,  \n
\delta \bar{\mu} &=& \epsilon \partial \bar{\mu}+(\bar{\mu}\partial
-\bar{\mu}^2\bar{\partial})\epsilon, \n
\delta f &=& \epsilon \partial f, \quad \quad \delta \bar{f} =
\epsilon \partial \bar{f}.
\label{8}
\eeqa
Because $f$ and $\bar{f}$ are not local functions of $\mu$ and $\bar{\mu}$,
$S_{ind}$ is non-local.

We note that in the gauge (\ref{1}) $\mu$, $\bar{\mu}$ and $\phi$ are
expressed in terms of $\sqrt{\gamma}$ as
\beq
\mu=\bar{\mu}=(1-\sqrt{\gamma})/(1+\sqrt{\gamma}), \quad \phi=2 \ln \{ (1+
\sqrt{\gamma})/2 \}.
\label{9}
\eeq
Conversely, $\gamma$ is a function of $f$ and $\bar{f}$, and by using
(\ref{4}) we obtain
\beq
\sqrt{\gamma}=-i\partial_1 f/\partial_0 f=i \partial _1 \bar{f}/\partial_0
                          \bar{f}.
\label{10}
\eeq
Thus $f$ and $\bar{f}$ are not independent.
In the case of pure gravity the action $S$ consists of the cosmological term
and the induced action coming from the Faddeev-Popov determinant.  Here the
ghost coordinates are integrated out.  By substituting (\ref{9}) into
(\ref{5}), we obtain
\beqa
S &=& \frac{\kappa}{4\pi} \int d^2x [ (3+3\sqrt{\gamma}+\gamma)
(1+\sqrt{\gamma})^{-3} (\partial_0
\sqrt{\gamma})^2 -(1+\sqrt{\gamma})^{-3}(\partial_1 \sqrt{\gamma})^2 \n
& & +\lfrac{1}{2} \{ \frac{\partial^2f \partial \bar{\partial} f-{\partial}^3f
\bar{\partial}f }
{(\partial f)^2} + \mbox{c.c.} \} +4\lambda \sqrt{\gamma} ],
\label{11}
\eeqa
where $S_{ind}$ is multiplied by some constant, $\kappa /(4\pi)$,
which is proportional to the conformal anomaly.

We would like to reduce the problem of two-dimensional field theory to that of
quantum mechanics.  This will be done by using the fact that the stress-energy
tensor must vanish.  We will first modify the gauge condition of (\ref{1}) to
\beq
g_{00} = 1+ h_{00}, \quad \quad g_{01} = h_{01},
\eeq
where $h_{00}$ and $h_{01}$ are some fixed functions.  The partition
 function $Z$, which is reparametrization invariant, should be independent of
$h_{00}$ and $h_{01}$ and we have the condition \cite{kpz}
\beq
\left. \frac{\delta Z}{\delta h_{00}(x)} \right | _{h_{00}=h_{01}=0} =
 \left.  \frac{\delta Z}{\delta h_{01}(x)} \right | _{h_{00}=h_{01}=0} =0.
\eeq
This is equivalent to the vanishing of the stress-energy tensor
$T_{00}=T_{01}=0$.
By differentiating $Z$ w.r.t. $h_{00}$ and $h_{01}$ arbitrary times we
conclude that correlation functions of arbitrary number of $T_{00}$'s and
$T_{01}$'s are zero.
\footnote{Contact terms may appear because in the action (\ref{5}) the time
components $g^{00}$ and $g^{01}$ are not simply the Lagrange multipliers of
the constraints $T_{00}=T_{01}=0$.  We will neglect such contact terms.}
Therefore we can substitute the solution of $T_{00}=0$ and/or $T_{01}=0$ into
the action (\ref{11}).  In this letter we will solve only the constraint
$T_{01}=0$ \footnote{It is known that the constraint $T_{00}=0$ is not
simply satisfied.}  explicitly and use this constraint to reduce the model
(\ref{11}) to quantum mechanics.  Validity of this procedure will be
justified later by the result (\ref{Sred2}), which coincides with (\ref{S}).

The stress-energy tensor of the model (\ref{11}) is obtained by varying
(\ref{5}) w.r.t. $g^{\mu \nu}$ and we have
\beqa
\frac{4\pi}{\kappa} T_{00} &=& \han (F+\bar{F}) + \frac{1}{\sqrt{\gamma}}
 \partial_0^2 \sqrt{\gamma}-2\lambda,
\label{16}   \\
\frac{4\pi}{\kappa} T_{01} &=& \lfrac{i}{2} \sqrt{\gamma} (F-\bar{F}),
\label{17}   \\
\frac{4\pi}{\kappa} T_{11} &=& -\han \gamma (F+ \bar{F}) -2\lambda \gamma,
\label{18}
\eeqa
where
\beq
F= \{ f,x^0 \} = \partial_0^3 f/ \partial_0 f - \lfrac{3}{2} (\partial_0^2f /
 \partial_0 f)^2,  \quad \bar{F} = \{ \bar{f}, x^0 \}
\label{19}
\eeq
are Schwarzian derivatives.

The solution to the constraint $T_{01}=0$ turns out to be
\beqa
f(x^0,x^1) &=& A \, \, tanh [ \int_0^{x^0} \frac{ds}{\sqrt{\gamma_0(s)}} +i \,
\int_0^{x^1} d \sigma \sqrt{\gamma_1(\sigma)} ] + C ,
\label{20} \\
\gamma (x^0,x^1) &=& \gamma_0 (x^0) \gamma_1 (x^1),
\label{21}
\eeqa
where $\gamma_0(x^0)$ and $\gamma_1(x^1)$ are arbitrary functions of $x^0$ and
$x^1$, respectively, and $A$ and $C$ are arbitrary constants.
This will be proved below.  Before that, two remarks are in order.
First, $\gamma$ is determined from (\ref{20}) by using (\ref{10}) and there is
no ambiguity associated with multiplying $\gamma _0$ by a constant and
dividing $\gamma _1$ by the same constant.  Secondly, $f$ should be a
single-valued function on a cylinder.    Thus from (\ref{20}) we conclude
that $\gamma _1$ has to satisfy a condition
\beq
\int_0^{\pi} dx^1 \sqrt{\gamma _1(x^1)} = (m+1) \pi \quad
            (m=0,1,2, \ldots ),
\label{22}
\eeq
where $0 \leq x^1 \leq \pi$ is a cycle of the cylinder.  Although we can
transform $\gamma _1(x^1)$ into a constant $(m+1)^2$ by reparametrizations of
$x^1$, two $\gamma_1$'s corresponding to different values of $m$ cannot be
connected by reparametrizations.  Thus in order to compute a cylinder
amplitude, we have to do summation over $m$.

Let us now prove (\ref{20}-\ref{21}).  By using (\ref{10}) we can show the
following identities by direct calculations
\beqa
\partial_1 F -i \sqrt{\gamma} \partial_0F-2i \partial_0 \sqrt{\gamma}F
     &=& i \partial _0^3 \sqrt{\gamma},
\label{23} \\
\partial_1 \bar{F} + i\sqrt{\gamma} \partial_0 \bar{F} +2i \partial_0
\sqrt{\gamma} \bar{F} &=& -i \partial_0^3 \sqrt{\gamma}.
\label{24}
\eeqa
Because of (\ref{17}) the constraint $T_{01}=0$ means that $F= \bar{F}$,
and hence we can show from (\ref{23}) $+$ (\ref{24}) that $F$ is a function of
only $x^0$.  Then (\ref{23}) can be integrated w.r.t. $x^0$;
\beq
F(x^0)= \bar{F}(x^0) = -\frac{1}{\sqrt{\gamma}} \partial_0^2 \sqrt{\gamma} +
  \frac{1}{2\gamma} (\partial_0 \sqrt{\gamma})^2 - \frac{2}{\gamma} \gamma_1
(x^1),
\label{25}
\eeq
where $\gamma_1(x^1)$ is an arbitrary real function of $x^1$.  The non-linear
differential equation for $f$ obtained by combining (\ref{19}) and (\ref{25})
has a general solution of the form
\beq
f(x^0,x^1)=A(x^1) \, tanh [ \sqrt{\gamma_1(x^1)} \int_0^{x^0} \frac{ds}{\sqrt{
\gamma (s,x^1)}} + B(x^1) ] + C(x^1),
\label{26}
\eeq
where $A(x^1)$, $B(x^1)$ and $C(x^1)$ are arbitrary functions of $x^1$.
Finally by requiring that (\ref{26}) should satisfy (\ref{10}), we obtain
(\ref{21}) and find that $A$ and $C$ are constants and that
\beq
B(x^1)= i \int_0^{x^1} d \sigma \sqrt{\gamma_1 (\sigma)} .
\label{27}
\eeq

We will next reduce the model (\ref{11}) to quantum mechanics by using
the solution  (\ref{20}-\ref{22}) to the constraint $T_{01}=0$.
\footnote{
This procedure is subtle because $T_{01}$ and $T_{00}$ do
not  commute.  Direct substitution of (\ref{20}-\ref{22}) into (\ref{11})
does not yield the last term in (\ref{28}).
This term, however, should exist because it is identified as
the Casimir energy for the cylindrical configuration.
}
The simplest way to do this is to compute the Hamiltonian
\beq
H_{red}= \int_0^{\pi} dx^1 \, 2\, \sqrt{\gamma} \, T_{00}.
\eeq
By using (\ref{16}), (\ref{21}), (\ref{22}) and (\ref{25}), we obtain
\beq
H_{red} = \frac{\kappa}{2} (m+1) [ \frac{1}
{2\sqrt{\gamma_0}} (\frac{d}{dx^0}\sqrt{\gamma_0})^2-2\lambda \sqrt{\gamma_0}
-\frac{2}{\sqrt{\gamma_0}} ].
\label{29}
\eeq
Then the action which yields $H_{red}$ is found to be
\beq
S_{red} = \frac{\kappa}{2} (m+1) \int_0^{T} dx^0 [\frac{1}{2\sqrt{\gamma_0}}
(\frac{d}
{dx^0} \sqrt{\gamma_0})^2 + 2\lambda \sqrt{\gamma_0}
                    + \frac{2}{\sqrt{\gamma_0}} ],
\label{28}
\eeq
where $T$ is the proper time.
\footnote{It was suggested in \cite{Moore} without proof that the problem of
2d gravity reduces to that of quantum mechanics.  Classically $S_{red}$ is
related to the minisuperspace model \cite{Moore} in the conformal gauge by a
transformation $x^0 \rightarrow \tilde{x}^0$ with $dx^0/d
\tilde{x}^0 =\sqrt{\gamma_0}=exp \,( \phi /2)$.     }
The variation of $S_{red}$ w.r.t. $\gamma_0$ satisfies the following relation:
\beq
\delta S_{red}= \int d^2x \sqrt{\gamma} \tilde{T}_{11}
\delta (\frac{1}{\gamma}),
\eeq
where $\tilde{T}_{11}$ is obtained by substituting (\ref{25}) into (\ref{18}).
Let us define the length of a loop at time $x^0$ by
\beq
l(x^0)= \frac{1}{\pi} \int_0^{\pi} dx^1 \sqrt{\gamma (x)}=(m+1) \sqrt{\gamma_0
(x^0)}.
\eeq
Then the action (\ref{28}) is rewritten  in terms of $l$ as
\beq
S_{red} = \kappa \int_0^T dx^0 [ \frac{1}{4} l^{-1}(x^0) \, \{ \frac{d}{dx^0}
l(x^0) \} ^2 + \lambda \, l(x^0) + (m+1)^2 \, l^{-1}(x^0) ].
\label{Sred2}
\eeq
Note that $S_{red}$ coincides with (\ref{S}) up to constant rescalings of the
variables and that $a$ takes descrete values $\kappa \, (m+1)^2$.

In the conformal gauge the constant $\kappa$ is determined by invariance of
the  partition function under the transformation
$\phi(x) \rightarrow \phi(x)+\epsilon(x)$, $ \hat{g}_{\mu \nu}(x)
\rightarrow e^{-\epsilon(x)} \hat{g}_{\mu \nu}(x)$ \cite{dkd}.
In the light-cone gauge the SL(2) current algebra determines
$\kappa$ \cite{kpz}.      In the present case  an argument which determines
the value of $\kappa$ is not available.   Here we will set $\kappa=1$,
because this leads to a simple and interesting two-loop amplitude.
It is straightforward to consider the case of arbitrary $\kappa$.  Later we
will also consider the case $\kappa=\infty$.
We also have to take into account the possibility that the coefficient
$m+1$ in (\ref{28}) may be additively renormalized by a Jacobian associated
with a change of variables from $f(x^0,x^1)$ to $\gamma_0(x^0)$.
It turns out that the action (\ref{Sred2}) with $m+1$ shifted by $-\han$,
\beq
S_m = \kappa \int_0^T dx^0 [ \, \frac{1}{4 \, l(x^0)}(\frac{d}{dx^0}l(x^0))^2
+\lambda l(x^0)+ \frac{(m+\han)^2}{l(x^0)} \, ]
\quad (\kappa=1)
\label{30}
\eeq
leads to the two-loop amplitude derived in matrix models.

To quantize the model (\ref{30}) we will switch to the Minkowskian metric
temporarily.  The Hamiltonian $H_m$ derived from $S_m$ is given by
\beq
H_m= \Pi_l \, l \, \Pi_l + (m+\han)^2 l^{-1} + \lambda l,
\label{32}
\eeq
where $\Pi_l$ is the canonical momentum conjugate to $l$ and is replaced by
$-i \partial / \partial l $ upon quantization.  We would like to solve the
problem of energy eigenvalues
\beq
H_m \Psi (l) = E \Psi (l) .
\label{33}
\eeq
Rescaling the variable $l \rightarrow z=2\sqrt{\lambda} l$ and defining a new
function $\Phi(z)=e^{z/2} z^{-(m+\han)} \Psi(l)$, we can rewrite (\ref{33})
into a confluent hypergeometric differential equation
\beq
z\frac{d^2}{dz^2} \Phi + (2m+2-z) \frac{d}{dz} \Phi -(m+1-\frac{E}
{2\sqrt{\lambda}}) \Phi = 0.
\label{34}
\eeq
The wave function should be regular at both $l=0$ and $l=+\infty$.  Solving
(\ref{34}) with this boundary condition, we obtain the following orthonormal
wave functions and energy eigenvalues
\beqa
\Psi_n^{(m)}(l) &=& \sqrt{\frac{n!}{(n+2m+1)!}} (2\sqrt{\lambda})^{m+1}
e^{-\sqrt{\lambda} l} l^{m+\han} L_n^{(2m+1)}(2\sqrt{\lambda} \, l),
\label{35} \\
E_n^{(m)} &=& 2 \sqrt{\lambda} (n+m+1),
\label{36}
\eeqa
where $n=0,1,2,\ldots$ and $L_n^{(\alpha)}(z) $ is Laguerre polynomial
\cite{El}.

We now turn to the construction of the cylinder amplitude
\beq
A_m(l_1,l_2;T)=<l_2|e^{-TH_m} |l_1> = \sum_{n=0}^{\infty} \Psi_n^{(m)}(l_2)
e^{-E_n^{(m)}T} \Psi_n^{(m)}(l_1) .
\label{37}
\eeq
As mentioned above, however, we have to do summation over $m$ in order to
take  all inequivalent classes of $\gamma_1(x^1)$ into account.  The full
amplitude will be given by
\beq
A(l_1,l_2;T)= \sum_{m=0}^{\infty} w_m A_m(l_1,l_2;T),
\label{38}
\eeq
where $w_m$ is a weight factor.
Let us first compute $A_m(l_1,l_2;T)$.  By using (\ref{35}-\ref{37}) and a
formula \cite{El}
\beqa
& & \sum_{n=0}^{\infty} \frac{n!}{\Gamma (n+\alpha+1)} L_n^{(\alpha)}(x)
           L_n^{(\alpha)}(y) z^n \n
& & =(1-z)^{-1} exp \{ -z(1-z)^{-1}(x+y) \} (xyz)^{-\alpha /2}
      I_\alpha (2(1-z)^{-1} \sqrt{xyz} ), \, \, |z| < 1,
\eeqa
we obtain
\beqa
A_m(l_1,l_2;T) &=& \sqrt{\lambda} \, cosech(\sqrt{\lambda}T) \,
     exp[-\sqrt{\lambda}(l_1+l_2) \, coth (\sqrt{\lambda}T) ] \n
   & &  \cdot I_{2m+1}( 2\sqrt{\lambda} \sqrt{l_1 l_2} \,
                cosech \{ \sqrt{\lambda}T \}),
\label{39}
\eeqa
where $I_{\nu}(z)$ and $K_{\nu}(z)$ in (\ref{40}) below are modified Bessel
functions.
To compare our result with that of matrix models, (\ref{39}) should be
integrated over $T$ from $0$ to $+\infty$.  By using a formula in \cite{Grad}
we have
\beq
G_m(l_1,l_2)= \int_0^{\infty} dT A_m(l_1,l_2;T)= K_{m+\han}
                (\sqrt{\lambda}l_a) I_{m+\han}(\sqrt{\lambda}l_b),
\label{40}
\eeq
where $a=1, b=2$ if $l_1 > l_2$ and $a=2, b=1$ otherwise.
This result was
in fact expected, because $G_m(l_1,l_2)$ is the solution of
\beq
H_m(l_1,\partial_{l_1}) G_m(l_1,l_2) = \delta(l_1-l_2) .
\label{41}
\eeq

Let us now consider the full amplitude (\ref{38}).
We do not have a definite argument to determine $w_m$.
If we set for instance
\beq
w_m=(-1)^m(2m+1),
\label{W1}
\eeq
then  by (\ref{38}) and
(\ref{40}) we obtain
\beqa
\int_0^{\infty} dT A(l_1,l_2;T) &=& \sum_{m=0}^{\infty} (-1)^m (2m+1)
\int_0^{\infty} dT A_m(l_1,l_2;T) \n
&=& \sum_{m=0}^{\infty} (-1)^m (2m+1) K_{m+\han}(\sqrt{\lambda}l_a)
I_{m+\han}(\sqrt{\lambda}l_b).
\label{42}
\eeqa
Due to the following Gegenbauer's addition formula \cite{El} \cite{Moore},
\beqa
K_{\nu}(R)R^{-\nu} &=& 2^{\nu} \Gamma (\nu) \sum_{n=0}^{\infty} (\nu+n)
C_n^{1/2}(cos \, \theta) K_{\nu+n}(z) I_{\nu+n}(\zeta) (z \zeta)^{-\nu}, \n
R &=& (z^2+ \zeta^2-2z \zeta cos \, \theta )^{1/2}, \quad | \zeta | < |z|,
\quad \nu \neq 0, -1, -2, \ldots ,
\label{Geg}
\eeqa
(\ref{42}) can be summed to yield
\beq
\int_0^{\infty} dT A(l_1,l_2;T)= \frac{\sqrt{l_1l_2}}{l_1+l_2}
    e^{-\sqrt{\lambda}(l_1+l_2)},
\label{43}
\eeq
where $C_n^{\nu}(z)$ is Gegenbauer polynomial.
This agrees with the result of matrix model calculations \cite{Moore}.
However, the two-loop amplitude derived in matrix models contains diagrams
with many branches, {\it i.e.} small universes, attached on the wall of the
cylinder.
In fact the analysis of \cite{kkmw} shows that space-time geometries with
infinitely many, very thin branches prevail in the cylinder amplitude.  On
such geometries the gauge (\ref{1}) cannot be chosen and hence the
calculations in the present letter and those in matrix models cannot be
compared.  Therefore we have to choose different $w_m$ and determination of
the value of $w_m$ is left to the future work.

Up to this point we have assumed that $\kappa=1$.
However the Faddeev-Popov determinant in the gauge (\ref{1}) is given by
\beq
\Delta_{FP}=[ \, det' \, (-\gamma^{-\han} \, \partial_0 \, \gamma^{\han} \,
\partial_0 \, ) \, \, det \, ( - \gamma^{-\frac{3}{2}} \, \partial_0 \, \gamma
^{\frac{3}{2}} \, \partial_0 \, ) ]^{\han},
\label{FP}
\eeq
where $det'$ implies the determinant for nonzero modes.  The operators
in (\ref{FP}) do not contain $\partial_1$ and hence (\ref{FP}) is ill-defined.
Therefore the value of $\kappa$ may be infinite.
In such a case we have to keep $\kappa$ in the above arguments.
The amplitude (\ref{39}) now reads
\beqa
A_m(l_1,l_2;T)
&=& \kappa \sqrt{\lambda} cosech(\sqrt{\lambda} T)
    exp [ \, - \kappa \sqrt{\lambda} (l_1+l_2) coth(\sqrt{\lambda} T)] \n
& & \cdot I_{ \kappa (2m+1)}(2 \kappa \sqrt{\lambda} \sqrt{l_1 l_2}
      cosech ( \sqrt{\lambda} T)).
\label{K1}
\eeqa
The $\kappa \rightarrow \infty$ limit of (\ref{K1}) depends on how $l_1$,
$l_2$ and $\lambda$ are rescaled.
If $l_1$, $l_2$ and $\lambda$ are kept fixed, then the $\kappa \rightarrow
\infty$ limit of (\ref{K1}) is given by
\beq
A_m(l_1,l_2;T) \sim \sqrt{\frac{\kappa (m+\han)}{2\pi l_1l_2 \, sinh(2\xi)}}
      \,   \,    exp(-\kappa U),
\eeq
where
\beq
U = \sqrt{\lambda}(l_1+l_2) coth (\sqrt{\lambda}T)-2(m+\han) \, coth \, \xi+
      2(m+\han) \, \xi
\eeq
and $\xi$ is defined by
\beq
sinh \, \xi = (m+\han) \, (\lambda l_1 l_2)^{-\han} \, sinh(\sqrt{\lambda}T).
\eeq
If we instead rescale $l_1$, $l_2$ and $1/ \lambda$ by an infinite amount,
that is, if we replace $l_j$ and $1/ \lambda$ by $\Lambda l_j$ and $\Lambda '
/ \lambda$, respectively, where $\Lambda$ and $\Lambda '$ go to $\infty$ as
$\kappa \rightarrow \infty$, then we have simply
\beq
\lim_{\kappa \rightarrow \infty} A_m \propto \delta (l_1-l_2).
\eeq
This means that the Hamiltonian for loop propagation is zero.

To recapitulate, in the pure gravity theory on a cylinder the
quantum-mechanical action (\ref{30}) for the length $l(x^0)$ of a loop was
derived by starting from the gravitationally induced action in the
proper-time gauge (\ref{1}) and solving the gauge constraint $T_{01}=0$.
Then the two-loop amplitude $A_m(l_1,l_2;T)$ (\ref{39}) was obtained by
the quantization of this model.

An interesting point is that the summation over $m$ in (\ref{38}) is a natural
consequence of the requirement of integration over all inequivalent classes of
$\gamma_1(x^1)$.  These degrees of freedom $m=0,1,2,\ldots$ are the modes
associated with the cycle of the cylinder.
It will be important to further clarify the meaning of these winding modes.
Let us also note that the amplitude (\ref{37}) satisfies the composition law
\beq
\int_0^{\infty} dl_2 A_m(l_1,l_2;T) A_m(l_2,l_3;T')=A_m(l_1,l_3;T+T'),
\label{44}
\eeq
while the full amplitude (\ref{38}) fails to satisfy (\ref{44}) due to the
sum over $m$.

In this letter only the cylinder, {\it i.e.}, the propagator was studied.
As a next step, we would like to introduce the time $T$ and the quantum
number $m$ to each boundary of a disk and a three-loop vertex.
If this is achieved, higher genus amplitudes may be obtained by sewing the
boundaries of vertices.

\vspace{1cm}
The author thanks N.~Ishibashi, H.~Kawai, Y.~Okamoto and K.~Yoshida
for discussions.

\end{document}